\documentclass[english]{cccconf}
\usepackage[comma,numbers,square,sort&compress]{natbib}
\usepackage{epstopdf}
\usepackage{amsmath}
\usepackage{xcolor}

\begin{document}

\title{DC-DC Converters Optimization in Case of\\ Large Variation in the 
Load}

\author{Alexander Domyshev\aref{sei,poly},
        Elena Chistyakova\aref{poly,idstu},
        Aliona Dreglea\aref{poly,hit},
        Denis Sidorov\aref{sei},
        Fang Liu\aref{csu}}



\affiliation[sei]{Melentiev Energy Systems Institute, Siberian Branch, Russian Academy Sciences, Irkutsk, Russia
               \email{domyshev@isem.irk.ru, \, dsidorov@isem.irk.ru}}

\affiliation[hit] {Harbin Institute of Technology, Harbin, China
\email{adreglea@gmail.com}}

\affiliation[poly] {Irkutsk National Research Technical University, Irkutsk, Russia}
               
\affiliation[idstu]{Institute for System Dynamics and Control Theory,
Siberian Branch, Russian Academy Sciences, Irkutsk, Russia
        \email{elena.chistyakova@icc.ru}}

\affiliation[csu]{Department of Automation,
Central South University, Changsha, China
        \email{csuliufang@csu.edu.cn}}

\maketitle

\begin{abstract}
The method for controlling a DC-DC converter is proposed to ensures the high quality control at large fluctuations in load currents by using differential gain control coefficients and second derivative control. Various implementations of balancing the currents of a multiphase DC-DC converter are discussed, with a focus on achieving accurate current regulation without introducing additional delay in the control system. Stochastic particle swarm optimization method is used to find optimal values of the PID controller parameters.  An automatic constraint-handling in optimization are also discussed as relevant techniques in the field.
\end{abstract}


\footnotetext{This work was in part supported by the Ministry of Science and Education of the Russian Federation FZZS-2024-0003 and in part by the National Foreign Experts Program of China (Grant No.~DL2023161002L).}

\section{Introduction}

\begin{figure}[h]
    \centering
    \includegraphics[scale=0.242]{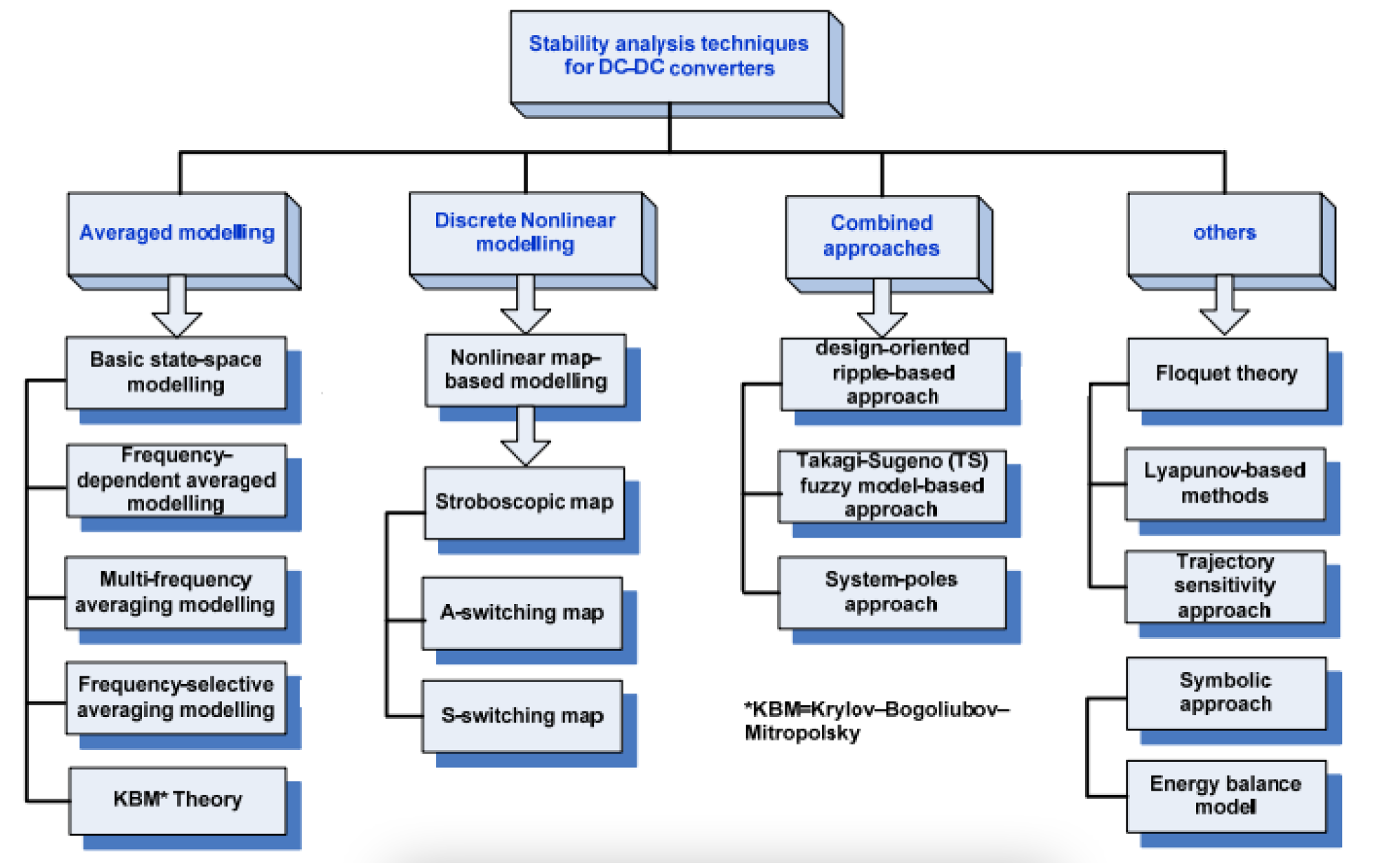}
    \caption{Classification of stability analysis methods \cite{thesis}}
    \label{fig:shema}
\end{figure}

DC-DC converters are essential electronic circuits that play a critical role in modern power management systems. Their primary function is to convert the voltage of a direct current (DC) source from one level to another, ensuring stable and efficient power delivery to various electronic devices and systems.
 In applications where input voltage levels can fluctuate due to factors such as battery discharging over time or changes in load conditions, DC-DC converters maintain a constant output voltage, providing reliable power to the system's components. DC-DC converters come in various topologies and configurations, catering to a broad range of applications and power requirements. In this paper we consider DC-DC control problem in case of big and fast load changes.
 We propose a method of controlling a DC-DC converter that provides high quality control at large fluctuations in load currents. The required speed of the converter duty cycle change is provided by sufficiently large values of the control coefficient of the differential gain. Stability of control and absence of overshooting is achieved by implementing compensating control by the second derivative of voltage. Due to the control link on the second derivative, the effect of predicting the transition to the stability border is provided.  There are various methods available for stability analysis techniques for DC-DC converters, as shown in Fig.~\ref{fig:shema}.

Converter designers frequently employ the state-space averaging technique to assess the stability and dynamic characteristics of power converters. This approach involves linearizing the actual nonlinear system around a steady-state operating point to produce a linear model. While this method provides a straightforward and precise model at a slower timescale, it may not accurately predict nonlinear behavior at a faster timescale. Modelling the dynamic
behaviour of pulse width modulation (PWM) controlled DC–DC converters mostly based on the the frequency-selective averaging method \cite{lit1}. The main challenge here is to 
effectively address the fast-scale nonlinearities. The range of available methods here employs the Poincaré map method, 
Floquet theory, Takagi–Sugeno fuzzy model-based approach and other.
For the comprehensive review of contemporary bifurcation analysis for power converter circuits and systems readers may refer to \cite{bifrev} and review of the stability analysis methods for switching mode power converters is given in \cite{ieeerev}.


 \section{Problem statement and model description}
 
 
The idealized model is a set of parallel inductances. The output capacitances of the individual circuits of the DC-DC converter are replaced by a single equivalent capacitance.
Idealized switches are used as keys connecting the inverter circuits to the input voltage and shunt keys. The bypass key control signal is inverted with respect to the linear key control signal.
The switches are controlled by a single PWM, which receives a value of the signal duty cycle as input. The duty cycle is calculated from the output voltage reference and the control signal from the PID controller, which is a voltage correction with respect to the nominal voltage.
In order to minimize the total peak current from the power supply, the control circuits are switched uniformly after a fixed time, which depends on the cycle time and the number of parallel circuits of the converter.  The model of the converter is shown in Fig. 2.

\begin{figure}
    \centering
    \includegraphics[scale=0.24]{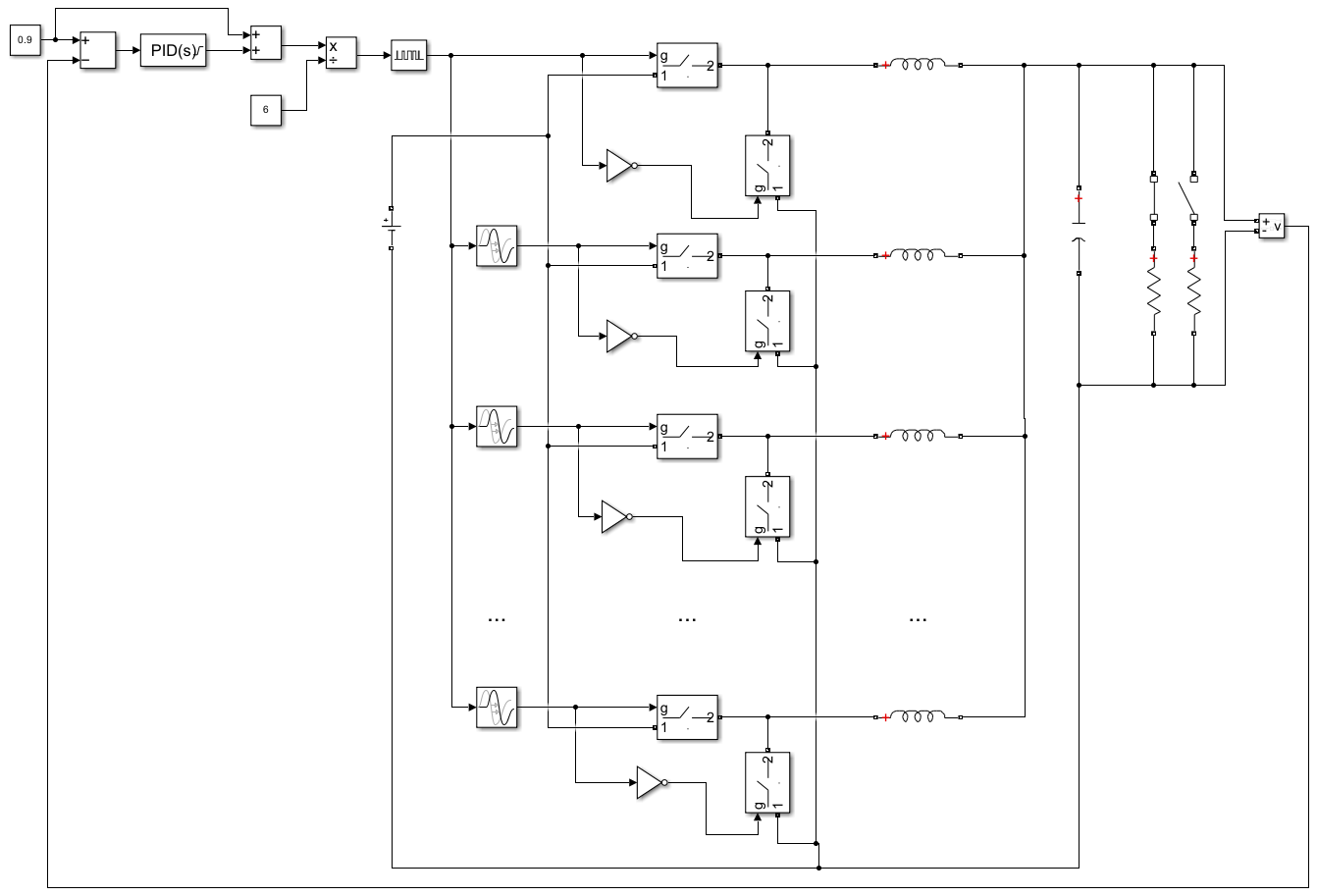}
    \caption{An idealized model of a DC-DC converter}
    \label{fig:enter-label}
\end{figure}

In the course of the research, the following solutions were developed to ensure the required quality of regulation
\begin{itemize}
    \item A regulator with a high derivative gain was used, providing the necessary speed of response to load changes.
\item 	To eliminate the overshoot in the second swing cycle and the subsequent instability, stabilization by the second derivative of the output voltage was applied.
\item 	To ensure that the output voltage returns to the reference value, the controller also uses integral and proportional gains.
\item An algebraic method is used to balance currents across phases, which does not add a delay during the controller’s response to disturbances.

\end{itemize}

\section{Mathematical Model}

The mathematical description of the DC-DC converter model with resistances in the capacitor and inductance circuit is the following system of differential, algebraic and integral equations:
\begin{equation}\label{modeq1}
\frac{dI_j}{dt}=\frac{1}{L_j}\bigl( \alpha_j(D_0)U_S-I_jR_L-U_O\bigr), \quad j=1,\ldots N,
\end{equation}
\begin{equation}\label{modeq2}
    \frac{dU_C}{dt}=\frac{1}{C}\Bigl(\sum\limits_{j=1}^N I_j-\frac{U_O}{R_{load}}\Bigr),
\end{equation}
\begin{equation}\label{modeq3}
    U_O=U_C+R_C\Bigl(\sum\limits_{j=1}^N I_j-\frac{U_O}{R_{load}}\Bigr),
\end{equation}
\begin{equation}\label{modeq4}
 T_d\frac{dU_{ad}}{dt}+U_{ad}=K_d\frac{de}{dt},   
\end{equation}
\begin{equation}\label{modeq5}
    U_{ai}=K_i\int\limits_{t_0}^te(\tau)d\tau,
\end{equation}
\begin{equation}\label{modeq6}
    T_{dd}\frac{dU_{dd}}{dt}+U_{dd}=K_{dd}\frac{d^2e}{dt^2},
\end{equation}
\begin{equation}\label{modeq7}
    U_a=U_{ad}+U_{ai}+K_pe+U_{dd},
\end{equation}
\begin{equation}\label{modeq8}
    e=U_{ref}-U_0,
\end{equation}
\begin{equation}\label{modeq9}
    D_0=\bigl(U_{ref}+U_a\bigr)/U_S,
\end{equation}
\begin{equation}\label{modeq10}
\alpha_j=
    \left\{ 
    \begin{array}{l}
         0, \quad \mbox{if} \quad (t-\Delta t_j)\, mod \, T>D_0T, \\
        1, \quad \mbox{if} \quad (t-\Delta t_j) \, mod \, T\leq D_0T,
    \end{array} 
    \right.
\end{equation}
In the control cycle, the second derivative of the voltage is calculated using the following expression
\begin{equation}\label{modeq11}
    \frac{d^2U_C}{dt^2}=\frac{1}{C}\Bigl(\sum\limits_{j=1}^N-\frac{1}{R_{load}}\frac{dU_C}{dt}\Bigr).
\end{equation}
Eqs. (\ref{modeq1})-(\ref{modeq11}) use the following notations: $N$ is number of circuits (phases) of the DC-DC converter, $I_j$ is phase current, $U_o$ is output voltage (actual), $D_0$ is total duty cycle, $U_{ref}$ is the desired output voltage to be maintained, $U_s$ is source voltage, $U_{ad}$ is voltage correction by controller differential gain, $U_{ai}$ is voltage correction by controller integration gain, $U_a$ is total voltage correction, $L_j$ is inductance in each circuit (inductances are assumed to be the same), $R_{load}$ is  load resistance (decreases abruptly -- once during the modeling process,the rate of change of resistance is equal to the current $dR_{load}=-0.9*2000\Sigma/us$, $R_C$ is total series capacitance resistance (ESR), $\Delta t_j$ is deviation of the control start time for each circuit (phase) of the DC-DC converter (uniform filling of the control cycle pulses for the entire DC-DC converter is assumed, so $\Delta t_j=T/N$. During the control process, the duration of the pulses changes, but their relative position along the time axis remains constant), $T$ is control cycle period, $T_d$ is inertia constant of the differential gain, $T_{dd}$ is inertia constant for the second order  derivative gain, $K_d$ is differential gain, $K_p$ is proportional gain, $K_i$ is integration gain, $K_{dd}$
is the correction factor for the second derivative. The controller constants to be optimized are $K_p$, $K_d$, $K_{dd}$, $K_i$, $T_d$, and $T_{dd}$. 

    
\section{Stability analysis}

Using elementary transformations,  Eqs. (\ref{modeq1})--(\ref{modeq11}) can be written as a following matrix equation:
\begin{equation}\label{smallsystem} 
{{d}\over{dt}}\begin{pmatrix}{\cal I}\cr  e\end{pmatrix}=\begin{pmatrix}B_1 & g_1\cr  -c &    G \end{pmatrix}\begin{pmatrix}{\cal I}\cr  e\end{pmatrix}+\begin{pmatrix}F_{1}\cr -F_{2}/r \end{pmatrix} 
\end{equation}
regulated by the remaining equations of the PID-controller 
\begin{equation}\label{PIDcon}
\begin{split}
 U_{a}=[K_{d}-K_{dd}b]e+K_{dd}{{de}\over{dt}}+\\
 +\int_0^t{\cal K}(t-\tau)e(\tau)d\tau + \varphi(t)   
 \end{split}
\end{equation}
and the switch (\ref{modeq9})-(\ref{modeq10}). Equations (\ref{smallsystem})-(\ref{PIDcon}) employ the following denotations:
$$
{\cal I}=\sum\limits_{j=1}^N I_j,
$$
$$
{\cal K}(t-\tau)=K_{i}-K_{d}a\exp[a(t-\tau)]+
$$
$$
+K_{dd}b^2\exp[b(t-\tau)],
$$
$$
\begin{gathered}
  \varphi(t)=\exp(at)U_{ad}(0)+\exp(bt)U_{dd}(0)-\\
  -K_{d}\exp(at)e(0)+ \\
  +K_{dd}b\exp(bt)e(0)-K_{dd}\exp(bt){{de}\over{dt}}(0),  
\end{gathered}
$$
$$
b=-1/T_{dd}, \ F_{1}={{1}\over{ L_j}}N[a(D_{0})U_{S}-U_{ref}], 
$$
$$
F_2=R_CF_1-\left[{{1}\over{CR_{load}}}U_{ref}+r{{d}\over{dt}}U_{ref}\right].
$$
The characteristic polynomial for system (\ref{smallsystem}) has the form
\begin{equation}
    \lambda^2 +(B_1+G)\lambda  + (B_1G+cg_1)=0,
\end{equation}
and, according to the Rauss-Hurwitz criterion, (\ref{smallsystem}) is stable if
\begin{equation}\label{RH}
(B_1+G)>0,\ (B_1G+cg_1)>0,   
\end{equation}
which is fulfilled automatically since 
$$
\begin{gathered}
B_1=-{{R_{L}}\over{L_j}},\ g_1={{1}\over{L_j}}N,\ -c={{1}\over{rC}}-{{R_{C}R_{L}}\over{r L_j}},\\
G=-\left[{{1}\over{rCR_{load}}}+{{R_{C}}\over{r L_j}}N_{f}\right].
\end{gathered}
$$
However, if we carry out the stability analysis using only the tools of the theory for differential equations, we are not able to provide an automatic algorithm for selecting parameters of the PID controller. The methods that we used so far do not take into account the discontinuity of 
the free term in the system caused by the switch (\ref{modeq10}). The concept of stability of the equilibrium solutions of a 
continuous, smooth system is well defined. However, for discontinuous systems,
 even the definition of solution is itself not straightforward.  Therefore, at this stage of research, we employ heuristic techniques for finding parameters of the PID controller.   

\section{Modeling Results}

\begin{figure}[h]
    \centering
    \includegraphics[scale=0.2]{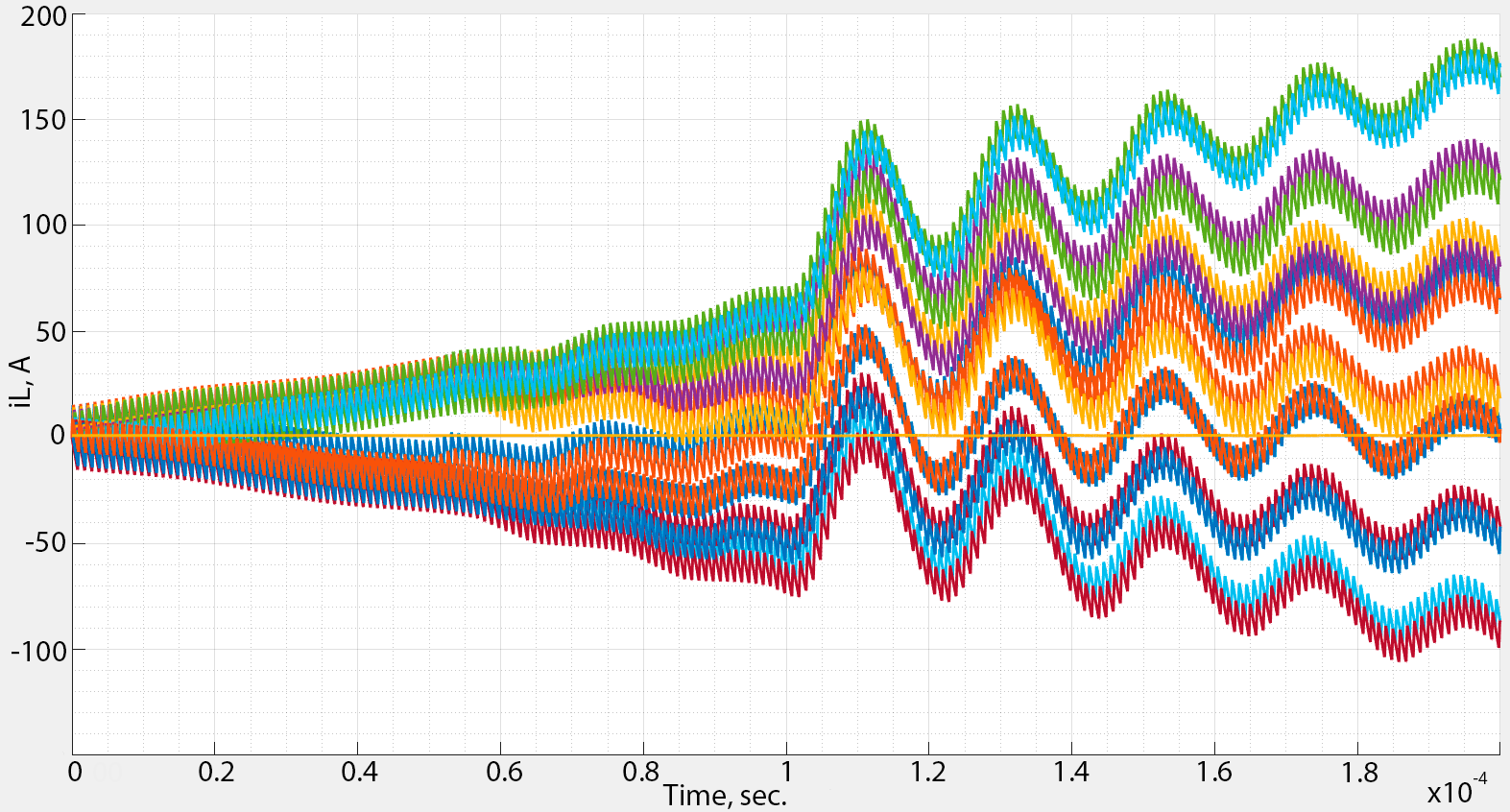}
    \caption{Voltage boosting started by voltage derivative threshold}
    \label{fig:currents_wo}
\end{figure}

The DC-DC converter regulator calculates the overall duty cycle of the multiphase converter based on the measurement of the output voltage.
To reduce current surges and ensure a smoother change in output voltage, the control cycle corresponding to the sampling frequency of the DC-DC converter is evenly filled with switching pulses in phases. In this case, with large duty cycle values, it is possible to overlap the state of several phases (several phases can be turned on simultaneously).
To model a control system, it is assumed that the duty cycle changes only at the beginning of each program (computational) cycle of the control system. The control cycle is assumed to be equal to the total sampling frequency of the DC-DC converter. Thus, there is some stair-stepping effect in the graph of the duty cycle changes during the control process (Fig \ref{fig:currents_wo}).

If the duty cycle is the same for all phases, then over time the currents may diverge across phases. This is due to the fact that the phases are switched on at different output voltages.
To ensure phase balancing, an algorithm has been developed to adjust the phase duty cycle according to the target duty cycle obtained at the output of the voltage control controller and phase current measurements.
In this case, a reliably working phase balancing algorithm can be performed using an arithmetic algorithm without the use of PID control. The advantage of this approach is that there is no need to adjust the control coefficients, which may differ for different operating modes of the regulator.

\begin{figure}[h]
    \centering
    \includegraphics[scale=0.2]{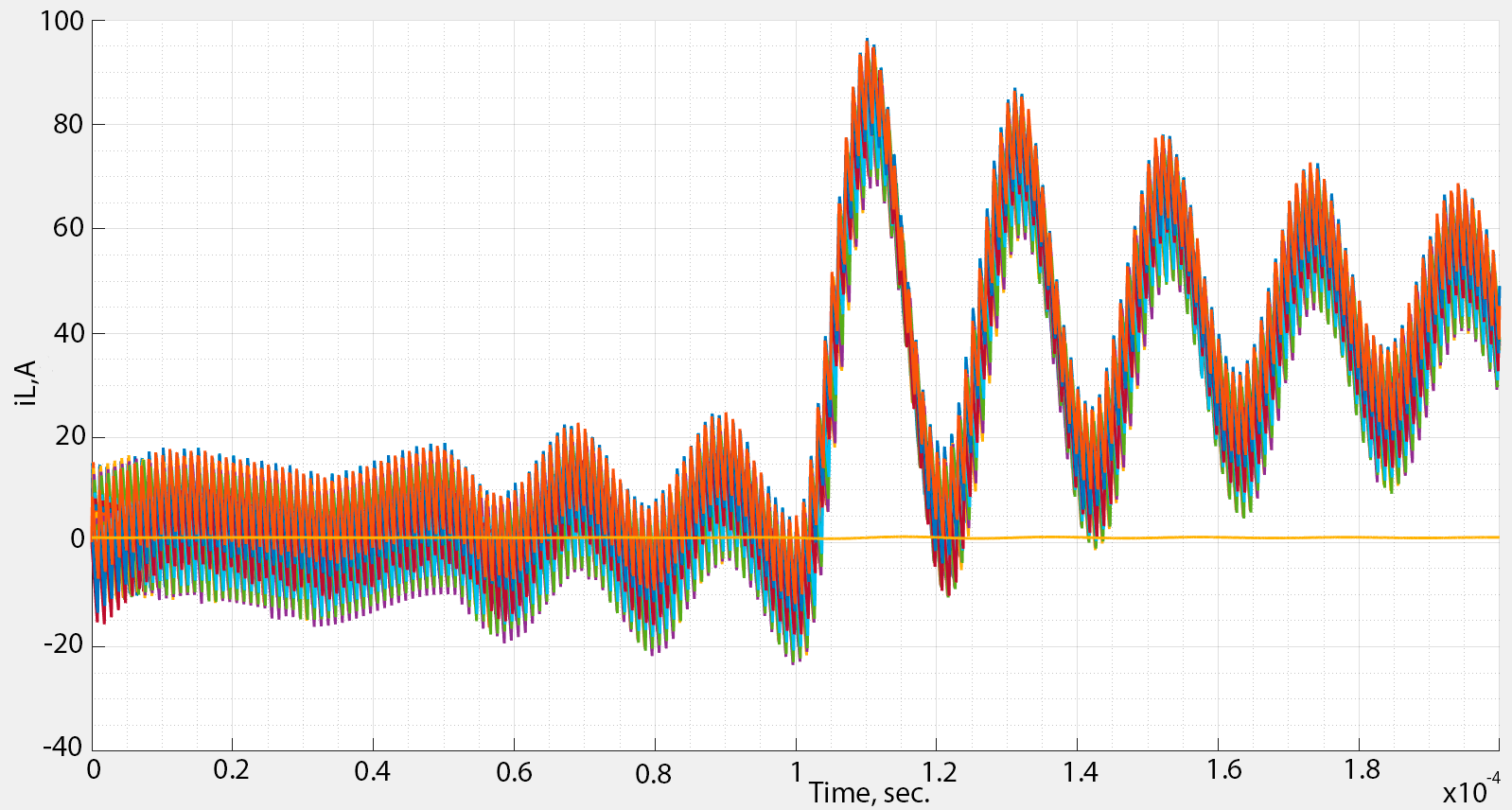}
    \caption{Phase currents with balancing algorithm}
    \label{fig:currents_withBA}
\end{figure}

\begin{figure}[h]
    \centering
    \includegraphics[scale=0.2]{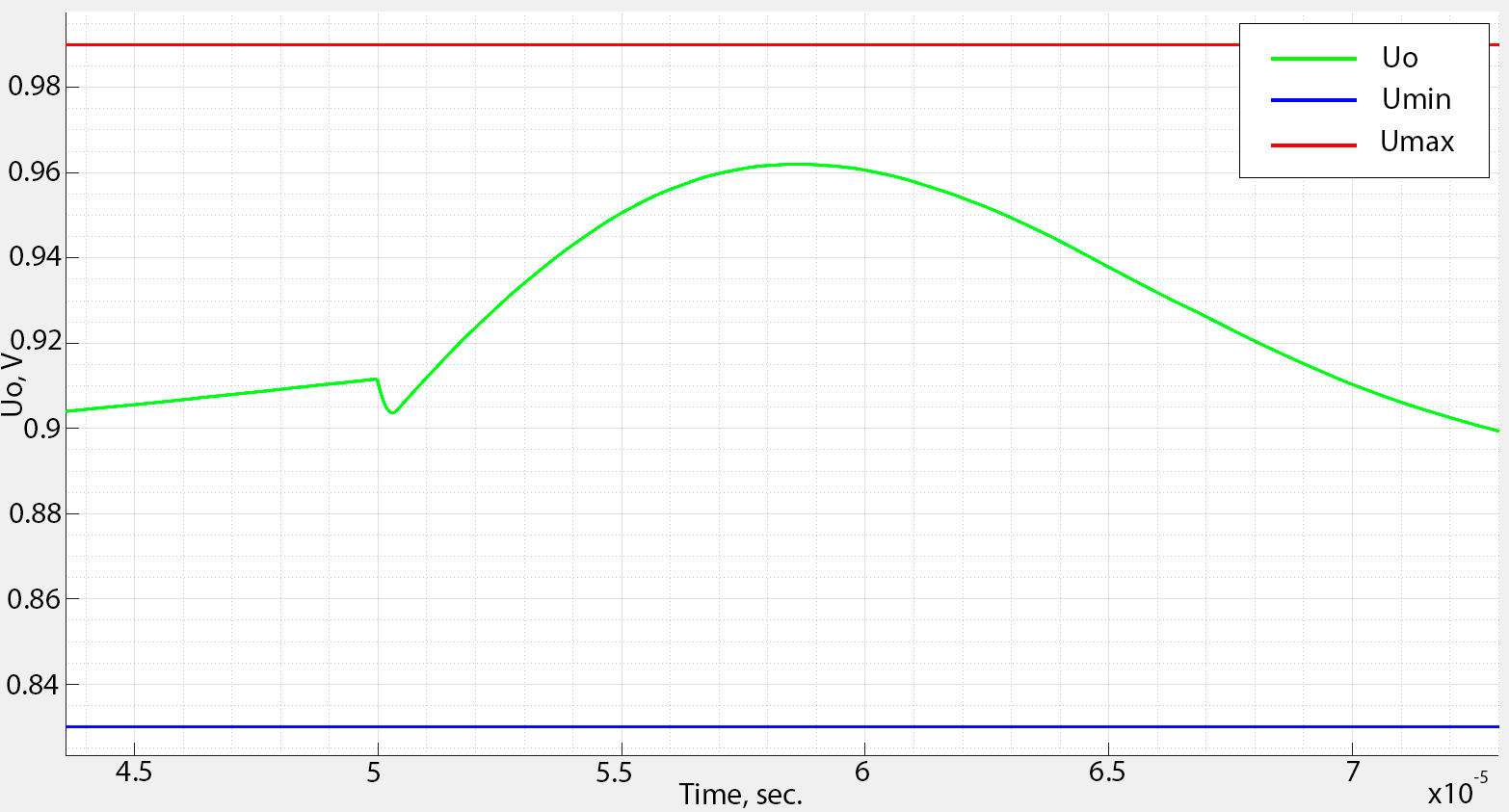}
    \caption{Voltage boosting started by voltage derivative threshold.}
    \label{fig:VB_1}
\end{figure}

The balancing algorithm:
\begin{enumerate}
    \item At each control cycle, moving averages of phase currents are calculated. In the controller, the calculation of the moving average $I_{{mean}_i}$ can be replaced by a low-pass filter.
    $$
    \hat{I}_{{mean}_i}=I_{{mean}_i}-\min_{i\in[1\ldots p]}I_{{mean}_i}.
    $$
    \item The duty cycle for each phase is calculated as
    $$
    D_i=\Bigl(1-\frac{\hat{I}_{{mean}_i}}{I_{sum}}\Bigr)D_0\frac{N}{N-1},
    $$
   where $I_{sum}=\sum\limits_{i\in[1\ldots p]}\hat{I}_{{mean}_i}$.
   \end{enumerate}
   Here  $D_0$ is the duty cycle value received from the controller; $N$ is the number of phases. 
   
   The duty cycle of each phase is calculated inversely proportional to the contribution of the phase current to the total current of all phases. To ensure balancing in the presence of negative phase currents, not the absolute value of the current is used, but its relative difference with respect to the minimum phase current ($I_{{mean}_i}$). Since in relative units the current of one phase is zero, the multiplier  $N/(N-1)$ is used.
   The change in currents by phase for the same transient process when using phase balancing algorithm is shown in Figure \ref{fig:currents_withBA}.


Experiments with voltage boost have shown that a quick response of the control system can be achieved by controlling the voltage derivative. We have a good quality of transient with manually adjusted action for the particular case (\ref{fig:VB_1}). The experiment shows that it is possible to keep voltage within the required range if we employ voltage derivative. 

Set a transfer function for the PID controller in the operator form as follows 
$$
F({\bf s})=K_p+\frac{K_i}{{\bf s}T_i}-\frac{{\bf s}K_d}{1+{\bf s}T_d}.
$$

\begin{figure}[h]
    \centering
    \includegraphics[scale=0.2]{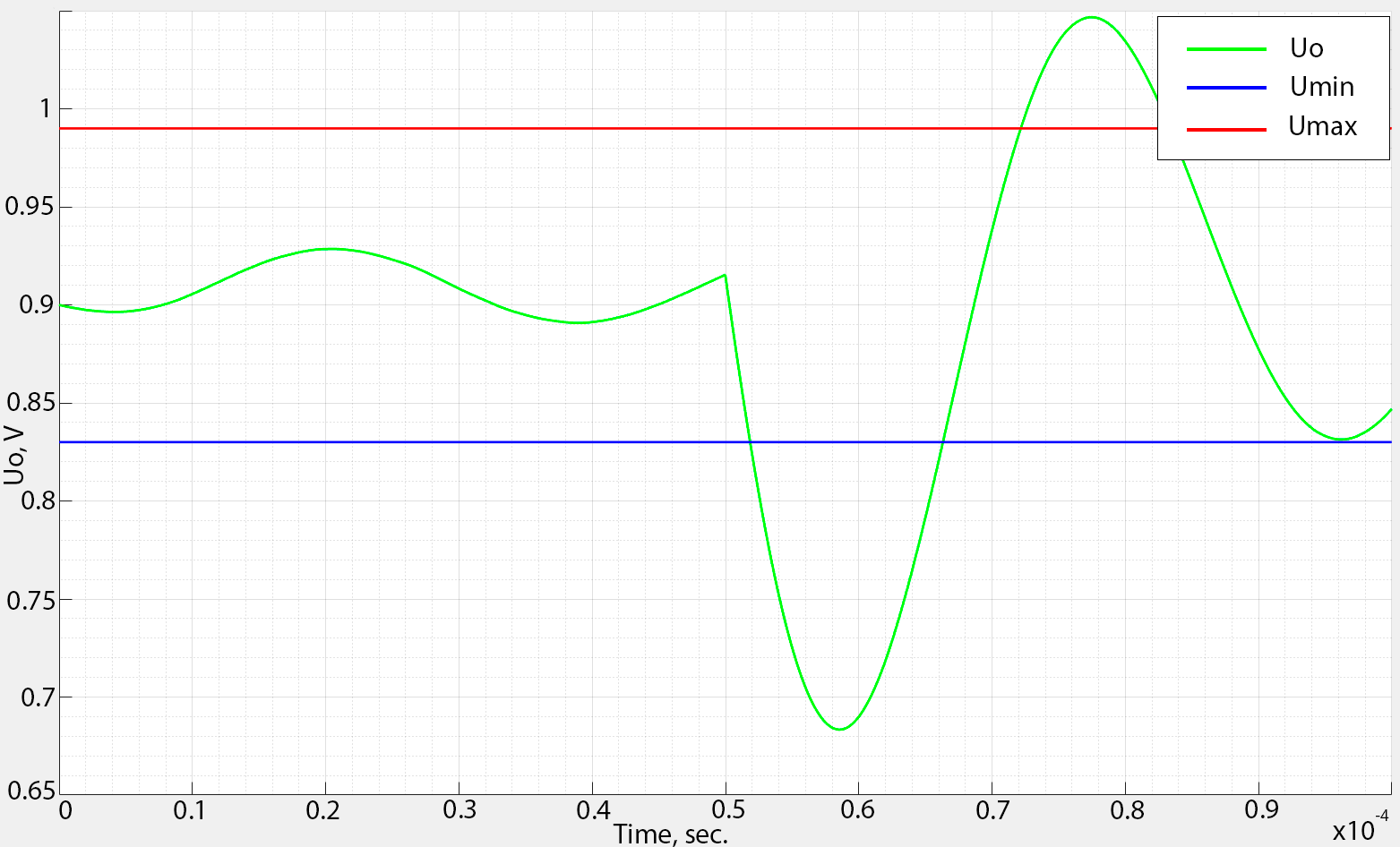}
    \caption{Voltage regulation with PID controller without derivative gain.}
    \label{fig:VB_2}
\end{figure}

Controlling a PID controller without using a differential gain does not provide the required response speed (Fig. \ref{fig:VB_2}).

\begin{figure}[h]
    \centering
    \includegraphics[scale=0.2]{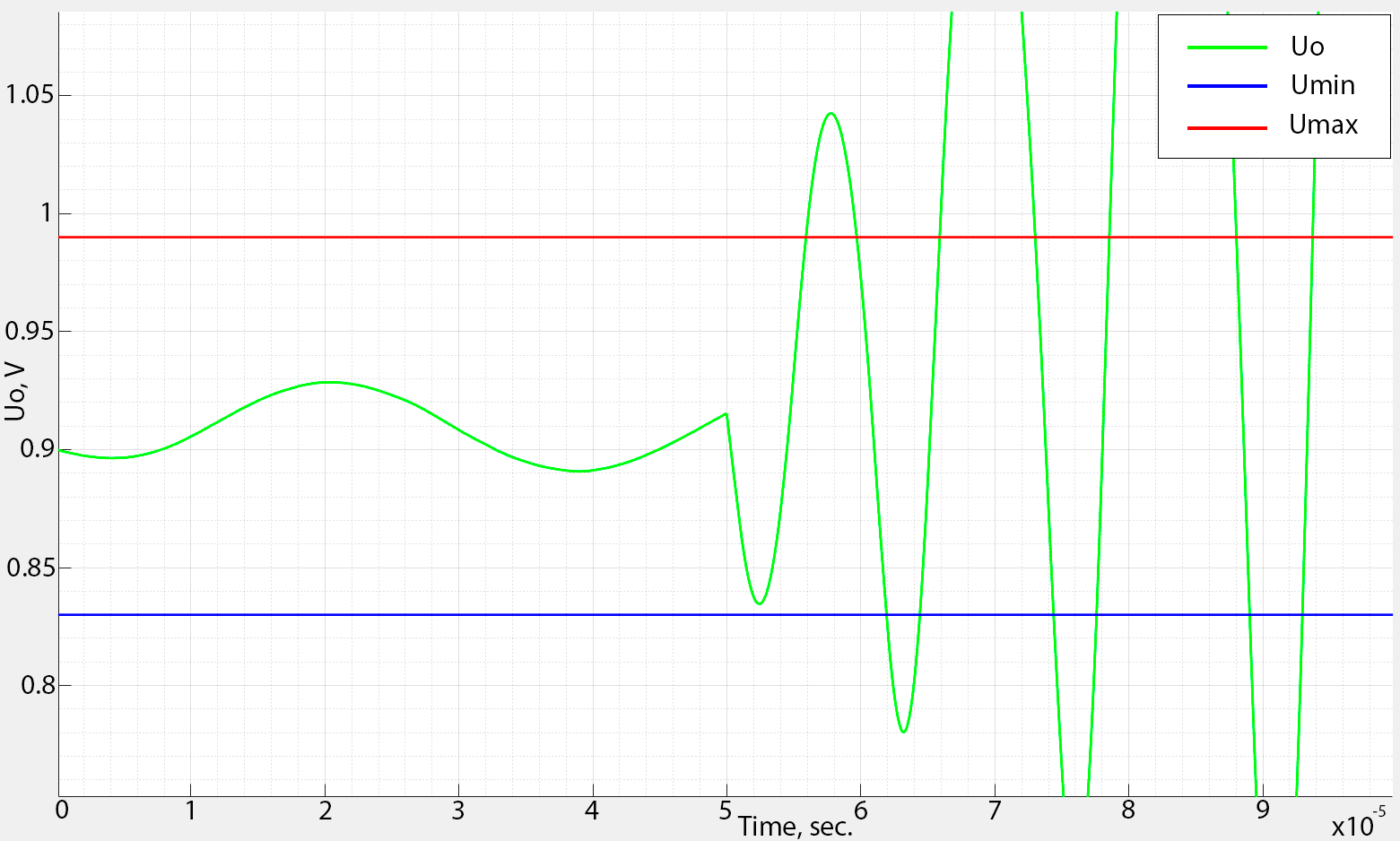}
    \caption{Voltage change when adding voltage derivative control}
    \label{fig:VB_3}
\end{figure}

Adding differential gain control provides the required speed of system response to disturbances. So, with a minimum value of $K_d$ and the time constant of the differential gain $T_d$, it is possible to keep the voltage above the lower permissible limit in the first oscillation cycle, but in the second and subsequent cycles the voltage goes beyond the permissible limits. Moreover, the system is oscillating (Fig. \ref{fig:VB_3}).

To provide corrective control that counteracts oscillations, a second derivative control gain has been added to the transfer function 
$$
F({\bf s})=K_p+\frac{K_i}{{\bf s}T_i}-\frac{{\bf s}K_d}{1+{\bf s}T_d}+\frac{{\bf s}^2K_{dd}}{1+{\bf s}T_{dd}}.
$$

\begin{figure}[h]
    \centering
    \includegraphics[scale=0.2]{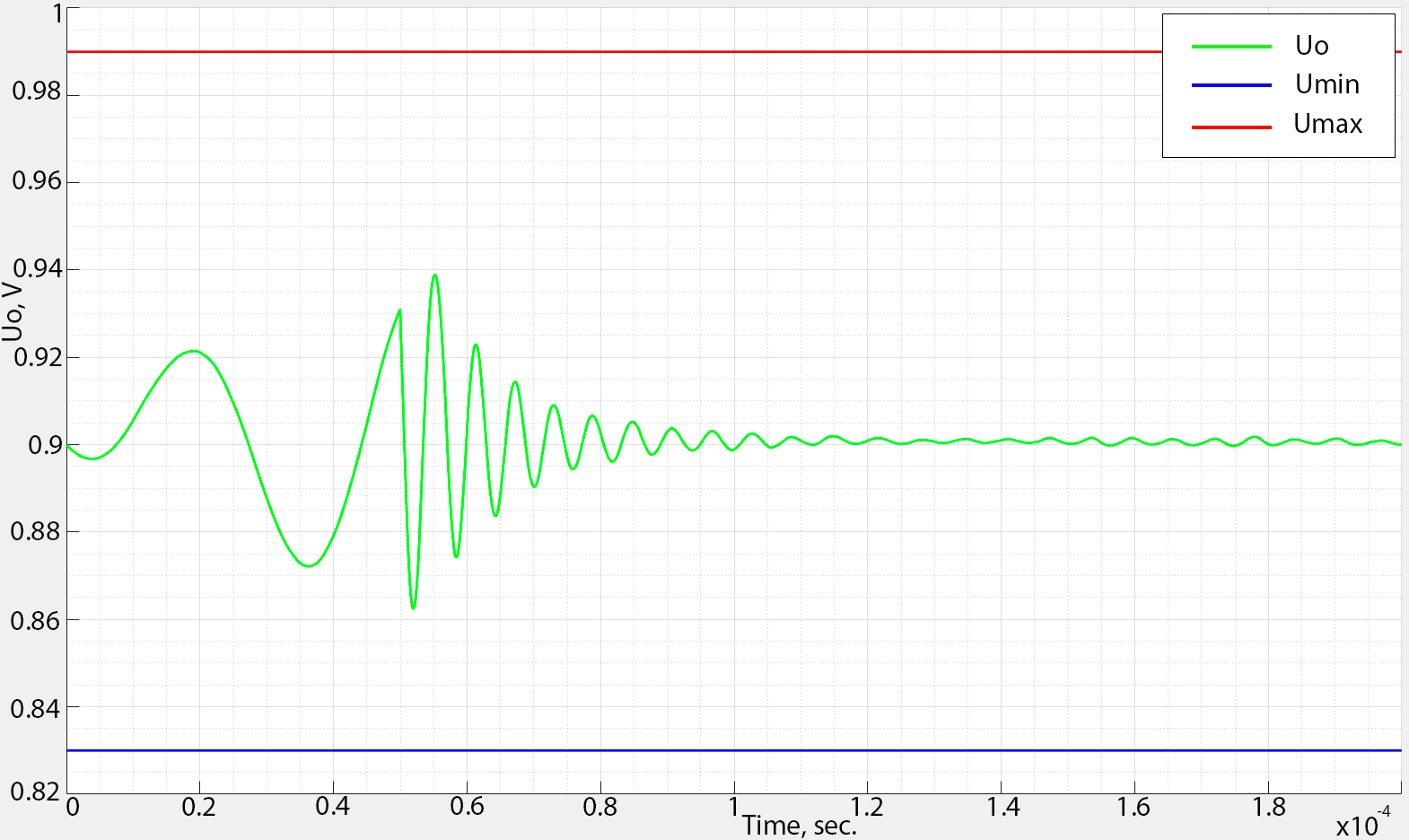}
    \caption{Changes in voltage when we enhanced the regulator by control via the second derivative of voltage}
    \label{fig:VB_4}
\end{figure}

The use of the second derivative in the control algorithm made it possible to keep the voltage in a given range and ensure universality of control (Fig. \ref{fig:VB_4}).  A change in the sign of the second derivative indicates that the system started to respond to control using the first derivative. Corrective control by the second derivative allows damping the regulation by the first derivative preventing overshoot.

\section{PSO Optimization}
To search for optimal values of the controller parameters, we used the stochastic optimization method – Particle Swarm Optimization (PSO) \cite{KennedyEberhart1948}-\cite{Pedersen2010}. PSO is a heuristic global optimization method, proposed originally by Kennedy and Eberhart in 1995  \cite{KennedyEberhart1948}. PSO performs searching via a swarm of particles that updates from iteration to iteration. To seek the optimal solution, each particle moves in the direction to its previously best position and the global best position in the swarm. PSO  makes few or no assumptions about the problem being optimized and can search very large spaces of candidate solutions.
Therefore, our choice of optimization method is stipulated by non-convexity of the objective function in the space of optimized variables. The PSO is easy to implement, has fast convergence, is capable of finding not only the local optimum but also the globally optimal solution and is easily scalable to handle problems of varying complexity and size. 
To optimize the controller parameters we used the following  objective function:
$$
F(X)=\ln(o(X)+\varepsilon)-\ln(\varepsilon)+\sigma(e(X)),
$$
s.t.
$$X\in[X_min;X_max]$$
where 
$X=[K_p, K_d, K_{dd}, K_i, T_{dd} ]$ is the control parameters vector.
$X_min$ and $X_max$ are constraints on control parameters obtained from experimental calculations.
$$
o(X)=
$$
$$
=\max(U_{min}+\varepsilon-\min_{\tau\in[t_0,t_e]}U_o(\tau); \max_{\tau\in[t_0,t_e]}U_o(\tau)-U_{max}-\varepsilon)
$$
is the outage 
of voltage over the allowable range with additional margin $\varepsilon=10^{-6}$. Standard deviation $e(.)$ is calculated for the entire modelling time range.
Optimal controller settings were tested at different load surge values (down to 1/10 of the maximum disturbance), as well as at different load increase rates (down to 1/10 of the maximum rate). The controller showed stable operation in all modes.


\section{Conclusions}

It is important to highlight that the proposed method for controlling a DC-DC converter ensures high quality control at large fluctuations in load currents. We achieved the required speed of the converter duty cycle by introducing sufficiently large values of the differential gain control coefficient. The second derivative of voltage was introduced as a compensating control to ensure stability and prevent overshooting. The control link on the second derivative helps to predict the transition to the stability border.
Conventional control algorithms for a multiphase DC-DC converter involve additional control of each phase duty cycle, when the deviation of phase current from the average of all phases is used as a control signal. Various implementations of balancing the currents of a multiphase DC-DC converter are possible. Sometimes the phase current control is performed by an additional PID controller following the common PID controller of the multiphase DC-DC converter. In other situations, additional correction at the output of the PID current controllers is passed to the input of a PID voltage regulator installed in each phase. The second option is more complex but provides more accurate current regulation. Both alternatives introduce additional delay in the control. In this paper, we proposed a simple variant of the current balancing realized as an arithmetic algorithm without PID control. It provides the necessary quality of balancing and uses only one PID controller in a multiphase DC-DC converter.

We used a stochastic PSO optimization method to find the optimal values of the PID controller parameters. The objective function was chosen so that it provides optimal control both at large load surges and in normal mode. Future work implies finding boundaries of the search area of optimal parameters through static stability analysis of the system of differential equations that models electric circuits incorporating a DC-DC converter and a controller. Such stability analysis should take into account the discontinuity of the right-hand part of the system of differential equations, which requires extra research.

\end{document}